\documentclass[manuscript]{aastex}

\newcommand\spitzer{\textit{Spitzer}}

\newcommand{\ltsimeq}{\raisebox{-0.6ex}{$\,\stackrel
        {\raisebox{-.2ex}{$\textstyle <$}}{\sim}\,$}}
\newcommand{\gtsimeq}{\raisebox{-0.6ex}{$\,\stackrel
        {\raisebox{-.2ex}{$\textstyle >$}}{\sim}\,$}}

\slugcomment{AJ ver \today}

\received{}
\revised{2011 Mar 18}
\accepted{}

\shortauthors{C.~E. Woodward et al.}
\shorttitle{C/2007 N3 (Lulin)}

\begin{document}

\title{Dust in Comet C/2007 N3 (Lulin)}

\author{Charles E. Woodward, Terry Jay Jones~\altaffilmark{1}, Bailey 
Brown, Erin Lee Ryan, Megan Krejny} 
\affil{Department of Astronomy, University of Minnesota \\ 116 Church 
Street S.E., Minneapolis, MN 55455 \\ chelsea@astro.umn.edu,
tjj@astro.umn.edu, baileybrown6@gmail.com, erinleeryan@gmail.com, 
krejny@astro.umn.edu} 

\author{Ludmilla Kolokolova, Michael S. Kelley}
\affil{Department of Astronomy, University of Maryland \\ College Park, MD 
20742-2421, ludmilla@astro.umd.edu, msk@astro.umd.edu} 

\author{David E. Harker}
\affil{Center for Astrophysics and Space Sciences, University of 
California, San Diego, 9500 Gilman Drive, La Jolla, CA 92093-0424, 
dharker@ucsd.edu} 

\author{Michael L. Sitko}
\affil{Department of Physics, University of Cincinnati, Cincinnati OH 
45221, and The Space Science Institute, 4750 Walnut Street, Suite 205
Boulder, CO 80301, fuse@amazon.net} 

\altaffiltext{1}{Visiting Astronomer at the Infrared Telescope Facility 
which is operated by the University of Hawaii under contract from the 
National Aeronautics and Space Administration.} 

\begin{abstract}


We report optical imaging, optical and near-infrared polarimetry, and 
\textit{Spitzer} mid-infrared spectroscopy of comet C/2007 N3 (Lulin). 
Polarimetric observations were obtained in $R$ (0.676~\micron) at 
phase angles from $0.44\degr$ to $21\degr$ with simultaneous 
observations in $H$ (1.65~\micron) at $4.0\degr$, exploring the 
negative branch in polarization. Comet C/2007 N3 (Lulin) shows a 
typical negative polarization in the optical
as well as a similar negative branch at near-infrared wavelengths. The 
10~\micron{} silicate feature is only weakly in emission and 
according to our thermal models, is consistent with emission from 
a mixture of silicate and carbon material. We argue that large, 
low-porosity (akin to Ballistic Particle Cluster Aggregates), 
rather absorbing aggregate dust particles best explain both the 
polarimetric and the mid-infrared spectral energy distribution.

\end{abstract}


\section{INTRODUCTION} \label{sec:intro}

Comet dust provides us with insight into two fundamental aspects of our 
solar system: (a)~the composition of both the solar system's initial 
ingredients and formation products; and, (b)~the spatial and temporal 
heterogeneity of the proto-planetary disk. Comet dust comae are optically 
thin, providing a unique laboratory for the study of preserved dust grains 
that observations of Kuiper Belt objects and primitive asteroids cannot 
provide. The compositions of both comet dynamical families, ecliptic 
comets (ECs) and nearly-isotropic comets (NICs), have been preserved in 
their interiors during their residences in the Scattered Disk and Oort 
Cloud. 


Knowledge of cometary materials enables better understanding of planetary 
and small body formation as well as evolutionary processes in the 
protoplanetary disk. The NASA space mission \textit{Stardust} was designed 
to bring back samples of comet dust from 81P/Wild 2. The terrestrial 
laboratory analysis of grain samples captured in the aerogel have helped 
us to understand the properties of comet refractory materials, revealing, 
among others, high-temperature crystalline materials \citep{flynn08}.  
In the preliminary analysis of \textit{Stardust} samples, $\simeq 65$\% 
of the aerogel tracks were `carrot-shaped,' indicative of large 
solid grains that upon inspection included micron- to tens of 
micron-sized forsterite (Mg-rich crystalline silicate) and FeS crystals 
\citep{burchell08, zolensky08}. A fraction, $\simeq 35$\%, of 
the \textit{Stardust} aerogel tracks were bulbous and attributed 
to porous aggregates which suffered significant heating 
upon impact with the aerogel \citep{roskosz08, leroux08}. However, 
analysis of microcrater impacts  in the foil suggest that $\simeq 85$\% 
are clustered with irregular, non-circular outlines likely
originating from composite, i.e., aggregate grains \citep{borg07}. In 
contrast to the \textit{Stardust} samples that were dominated by
weakly bound aggregates, which likely originated in the inner protoplanetary
disk, the interplanetary dust particles (IDPs) captured in Earth's 
stratosphere from comet 26P/Grigg-Skjellerup were all highly porous
\citep[e.g.,][]{naka08} and were not dominated by large single
mineral grains.


Grain thermal modeling of mid-infrared (MIR) remote sensing spectral 
energy distribution (SEDs) of comets constrains various coma grain 
properties including their: temperatures, size distributions ($n(a)da$), 
mass, dust production rate, and porosity \citep[e.g.,][]{harker02, woo99}. 
These grain properties must then be reconciled with those derived from 
analysis of the optical and near-infrared (NIR) polarimetry to produce a 
self-consistent description of the dust. 

The status of remote sensing of cometary dust was reviewed by 
\citet{kolv04}, who concluded that aggregate models are likely best at 
explaining photometric, polarimetric, and thermal properties of cometary 
dust. This model \citep{kol07} not only explained the phase dependence of 
brightness and polarization for comets, but also explained the existence 
of two classes of comets based on polarimetry by \citet{dob86} and 
\citet{Lev96}. \citet{Lis02} and \citet{sitko04} argue that there are two 
classes of comets based on the strength of the silicate feature in 
emission and the 10~\micron{} SED temperature excess over an equivalent
blackbody at the observed heliocentric distance ($r_{h}$) of the comet. 
Taxomomically, comets with strong silicate feature emission comprise
the same population of comets that show high polarization
(coma dust population dominated by small or very porous
particles), while those with weak silicate feature emission
are the same population that exhibit low polarization (coma dust
population dominated by large compact particles).
This correlation is consistent with 
aggregate dust models \citep{kol07}. Comparison of thermal model dust 
parameters used to interpret ground-based 
remote sensing observations of the SED arising from coma grains with the 
dust from the \textit{Stardust} sample return mission has helped to 
improve the aggregate model, making it not only qualitatively but also 
quantitatively compatible with the experimental data \citep{kolkim10}. 


The presence of the so-called negative branch in the phase dependence of the
polarization is an important observational characteristic of comet
polarization \citep{mishchenko10, kolv04}. Unlike Rayleigh
particles, which polarize light perpendicular (positive Stokes Q) to the
plane of scattering at all phase angles, for comets and asteroids the
polarization position angle flips by $90\degr$ (negative Stokes Q) at low
phase angles, $< 20$~\degr\, \citep[e.g.,][]{kel04}. Although
negative polarization can be reproduced by modeling comet dust as an
ensemble of polydisperse spherical particles \citep[e.g.,][]{mukai87}, 
\citet{kolv04} demonstrate that all other polarimetric, photometric,
and thermal IR characteristics of comet
dust require that comet partciles be aggregates of submicron monomers.
This aggregate ``structure'' was confirmed by the in situ
coma particle sampling of the \textit{Stardust} mission
\citep[e.g.,][]{flynn08} that showed that the majority ($\simeq 65$\%) 
of the returned
sample particles had an aggregate nature.  Reproduction of the
negative branch can be achieved in the aggregate model cases utilizing
large (thousands of monomers), aggregates \citep{kol07} or alternatively
in cases where the aggregates are transparent silicates \citep{zub09,
kim01}. However, the presence of aggregates made of transparent silicates
is hard to reconcile with the low geometric albedo of comet nuclei 
caused by a significant carbon content \citep{cam06}. Other ballistic
aggregate models \citep[e.g.,][]{lasue09, shen09, shen08} developed to
assess the effect of effective aggregate size, porosity, and composition
in the polarimetric behavior of dust in astrophysical environments need
to be carefully scrutinized for their applicability in describing dust in
comet coma. This assessment, which we discuss in detail in
\S\ref{sec:polmods}, is critical for interpreting the negative
polarization branch observed at small phase angles.

There are very few observations of comet polarization in the NIR at phase
angles below $20\degr$, and existing observations present a puzzle. The
unusual comet C/1995~O1 (Hale-Bopp), for example, showed no negative
branch at NIR wavelengths \citep{Jon00}, unlike the behavior of all
comets at optical wavelengths. Knowledge of the wavelength dependence of
the polarization at long wavelengths ($\lambda > 1.3~$\micron) through
the negative branch as well as NIR observations at a wide range
of phase angles should shed more light on the nature of the
aggregates that make up the majority of comet dust. These aggregates 
also need to reproduce the observed emission in the thermal infrared, 
especially in the 10~\micron{} spectral region where various minerals
produce broad resonances and narrow features.


The wavelength dependence of the polarization has only recently been 
extended out beyond $1~\micron$. By observing out to $2.2~\micron$, 
\citet{Jon00} and \citet{kel04} were able to show that in general, the 
polarization increases from the optical out to the NIR for most comets 
(red polarization color). This effect is in contrast to that observed in 
asteroids, where the polarization decreases with wavelength \citep[blue 
polarization color;][and references therein]{mas09}. Most NIR polarimetry 
observations of comets are made independent of the visual observations, 
requiring extrapolation between different phase angles in order to 
directly compare fractional polarization at visual and NIR wavelengths. 
This lack of simultaneity complicates determination of the polarization 
color. 

\begin{figure}
\epsscale{0.85}
\plotone{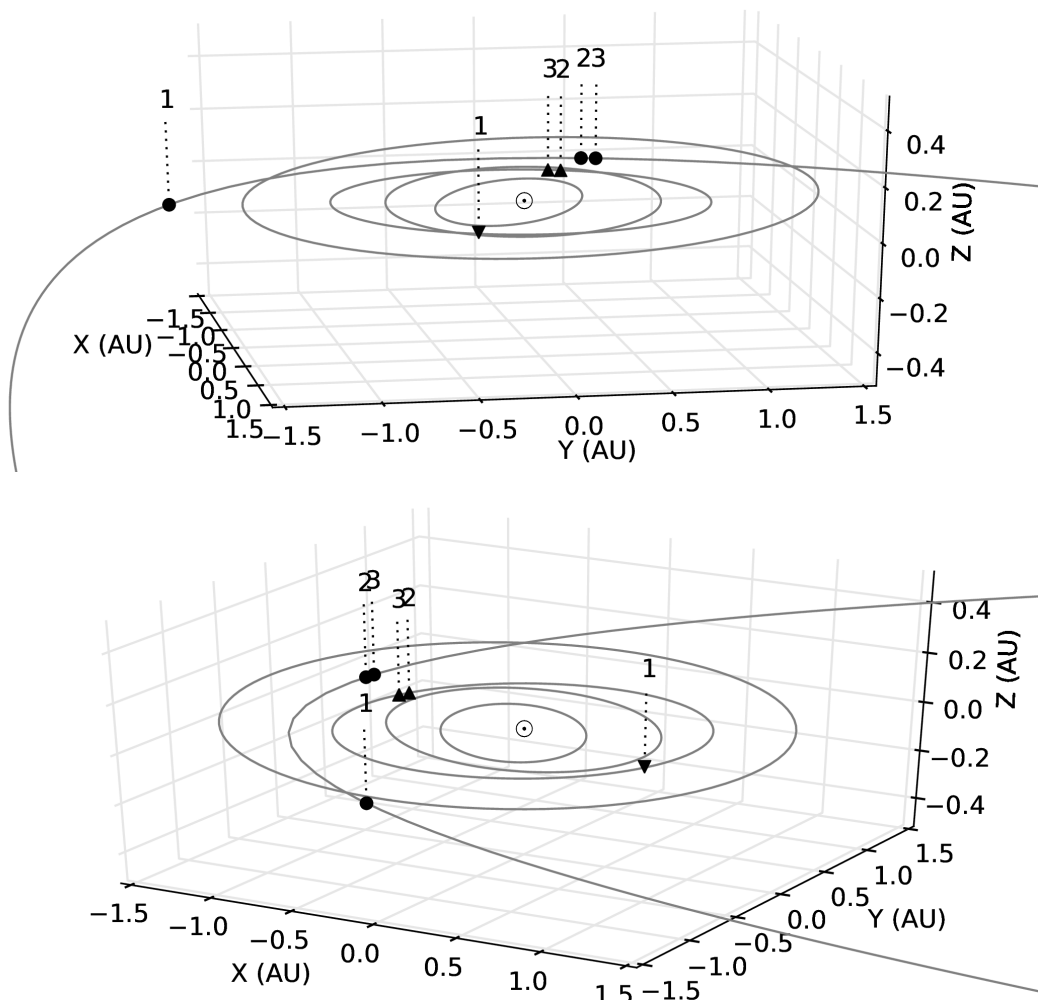}
\vspace{0.50cm}
\caption{Positions of the Earth (triangles) or \textit{Spitzer}
(upside-down triangle), and comet C/2007 N3 (Lulin) (filled
circles) on: (1) 2008 October 04~UT, (2) 2009 February 27~UT, and 
(3) 2009 March 03~UT depicted in heliocentric ecliptic 
coordinates.  The orbits of the terrestrial
planets, Mercury through Mars, and comet C/2007 N3 (Lulin) are drawn as 
gray lines, and the position of the sun indicated by the $\odot$. The
axis tick marks are in units of AU.
\label{fig:orbit_vis}}
\end{figure}

Comet C/2007 N3 (Lulin) provided an excellent opportunity to make 
simultaneous visual and NIR polarimetric observations of a comet at phase 
angles below $20\degr$. Comet C/2007 N3 (Lulin) is a NIC (Oort cloud) comet 
that passed less than $1\degr$ from zero phase angle at opposition
(see Fig.~\ref{fig:orbit_vis}). We 
obtained polarimetric observations in the optical red at phase angles from 
$0.44\degr$ to $21\degr$ with simultaneous observations in the NIR at 
$4.0\degr$. When combined with \textit{Spitzer} MIR spectroscopy, we can 
compare dust models that match the observed scattering and the observed 
emission properties of the dust in the coma. 

\section{OBSERVATIONS} \label{sec:obs}

\subsection{\textit{Optical Imaging}}\label{obs-bok}

Comet C/2007 N3 (Lulin) was observed on 2009 March 03~UT with the 2.3-m 
Bok Telescope at Kitt Peak National Observatory at heliocentric 
distance, $r_{h} = 1.45$~AU and a geocentric distance, $\Delta=0.49$ AU. 
The comet was at a phase angle $\alpha=17.77^{\circ}$. 

The images were obtained with the 90prime Camera \citep{williams04}, a 
prime focus imager built for the Bok Telescope. At the time of 
observation, the 90prime camera utilized a thinned back-illuminated CCD 
detector with 4064 $\times$ 4064 pixels with a pixel size of 
$15.0~\micron$. At prime focus the camera pixel scale is 
$0.45^{\prime\prime}$ which yields a field of view of 30.5$\times$30.5 
square-arcmin. The instrument was equipped with Cousins/Bessel system 
broadband $V$ and $I$ filters. 

\begin{figure}
\epsscale{1.0}
\plotone{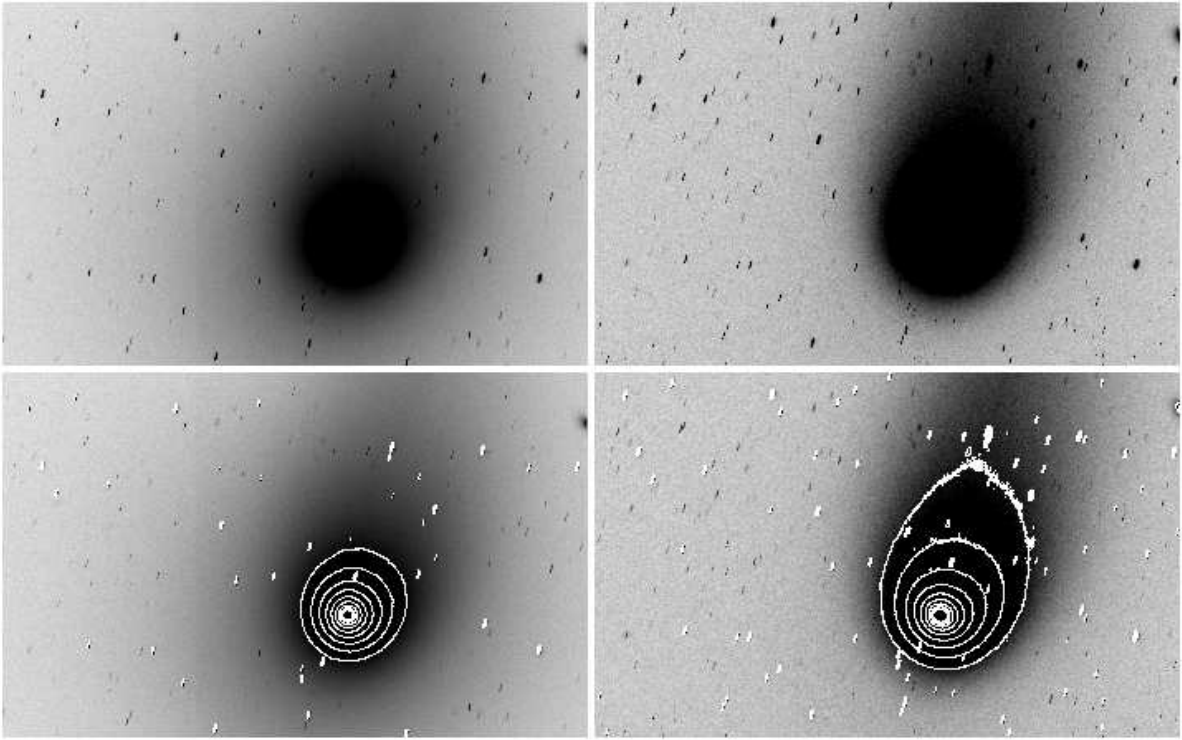}
\caption{(top) Greyscale $V$ band (left) and $I$ band (right) images 
of comet C/2007 N3 (Lulin) 
from the Bok 2.3-m Telescope. (bottom) same as above with isophotes
overlain on the image from 17 to 19 magnitudes per square-arcsecond at
0.25 magnitude per square-arcsecond intervals. The field of view
in each filter is 18.4 arcminutes x 14.3 arcminutes. The grey scale
intensity display table is proportional to the square of the counts.
\label{fig:bokVandI}}
\end{figure}

Multiple exposures (3 images of 120 seconds in $I$ band and 4 images in 
$V$ band; 1 of 60 seconds and 3 of 120 seconds) were obtained of the 
nucleus and coma of the comet with the telescope tracking at the 
non-sidereal rate corresponding to the predicted motion of the comet 
provided by JPL Horizons\footnote{http://ssd.jpl.nasa.gov/horizons.cgi} 
in an airmass range of 1.08 to 1.11. All images were corrected for 
overscan, bias and flat-fielding with standard IRAF\footnote{IRAF is 
distributed by the National Optical Astronomy Observatory, which is 
operated by the Association of Universities for Research in Astronomy 
(AURA) under cooperative agreement with the National Science Foundation.} 
routines. Due to the thinned nature of the 90prime detector, $I$ band data 
must be defringed. A master fringe file was created using the median of 41 
data frames from this night of observation and subtracted from the $I$ 
band data. The data was photometrically calibrated using Landolt standard 
fields PG0918, SA102, SA104, PG1323 \citep{landolt92} and the average 
nightly seeing was $1\farcs8$.  One 120 second exposure in each 
band is shown in Fig.~\ref{fig:bokVandI}. 

\subsection{\textit{Polarimetry}} \label{sec:pol}

The NIR observations were made in the $H (1.65~\micron)$ band using 
NSFCAM2 \citep{shure94} on the IRTF 3-m telescope at a plate scale of 
$0\farcs04$ per pixel, resulting in a field of view of $80\arcsec$ square. 
In polarimetry mode, NSFCAM2 utilizes a rotating half-wave plate at the 
entrance window of the camera and a cold wire grid polarizer in the second 
filter wheel. Details of the NSFCAM2 (+Polarimeter) observing technique 
and data reduction procedure are given in \citet{Jon00} and \citet{kel04}. 
The visual polarimetric observations were made using OptiPol \citep{Jon08} 
at the University of Minnesota Mt. Lemmon Observing Facility (MLOF) 
in a narrow band $R$ filter centered at $\lambda_{o} = 0.676$~\micron, 
with a bandwidth $\Delta\lambda = 0.04~\micron$. This filter was 
chosen to avoid contamination from unpolarized gas emission as 
much as possible. 

The results of our polarimetry, including the observing log, observation 
time, and phase angle are given in Table~\ref{tab:pol_tab}. Simultaneous 
polarimetry at $R$ and $H$ was obtained on 2009 January 27~UT. 

\subsection{\textit{Spitzer IRS}}\label{sec:obs-irs}

Comet C/2007 N3 (Lulin) was observed by \spitzer{} on 2008~October~04.3~UT 
($r_{h} = 1.90$~AU, a \spitzer{}-comet distance of 1.673~AU, and a phase 
angle of 32.5\degr) with the Infrared Spectrograph \citep[IRS;][]{houck04}  
as part of a larger Cycle~5 study assessing the water production and 
volatile production rates of comets (program identification number [PID] 
50335; PI: D.~E. Harker). The astronomical observation request (AOR) key 
for the dataset is \dataset[ADS/Sa.Spitzer#2587584]{2587584}, and the 
basic calibrated data (BCD) products processed with IRS reduction pipeline 
S18.60. The AOR for the short-wavelength, low-resolution SL1 ($7.4 - 
14.5$~\micron) data discussed here executed a $7\times 3$ spectral map 
(performed with no peak-up) yielding 21 spectra (6~sec $\times$ 2 cycles) 
with $1.87\arcsec\times10.0\arcsec$ \ steps (perpendicular $\times$ 
parallel to the long slit dimension). Background (shadow) observations 
were taken 34 hours later at the same celestial coordinates as the target 
spectra (AOR key 25988864), allowing the comet to move out of the spectral 
map field-of-view (FOV). Further analysis of the SL1 and LH spectra are 
discussed in \citet{woo11}. 
 
The spectra were reduced as follows.  The shadow observations were 
subtracted from the on-source observations, and the result was  assembled 
into a data cube using the CUBISM software \citep{smith07}, with bad 
pixels masked and all extended source calibrations applied.  The spectrum 
presented in this paper is extracted from an aperture 9\farcs25 $\times$ 
9\farcs25 in size, centered on the peak surface brightness of the comet. 

\section{DISCUSSION} \label{sec:disc}

\subsection{\textit{Optical Imaging}} \label{sec:dopt}

\begin{figure}
\epsscale{1.0}
\plotone{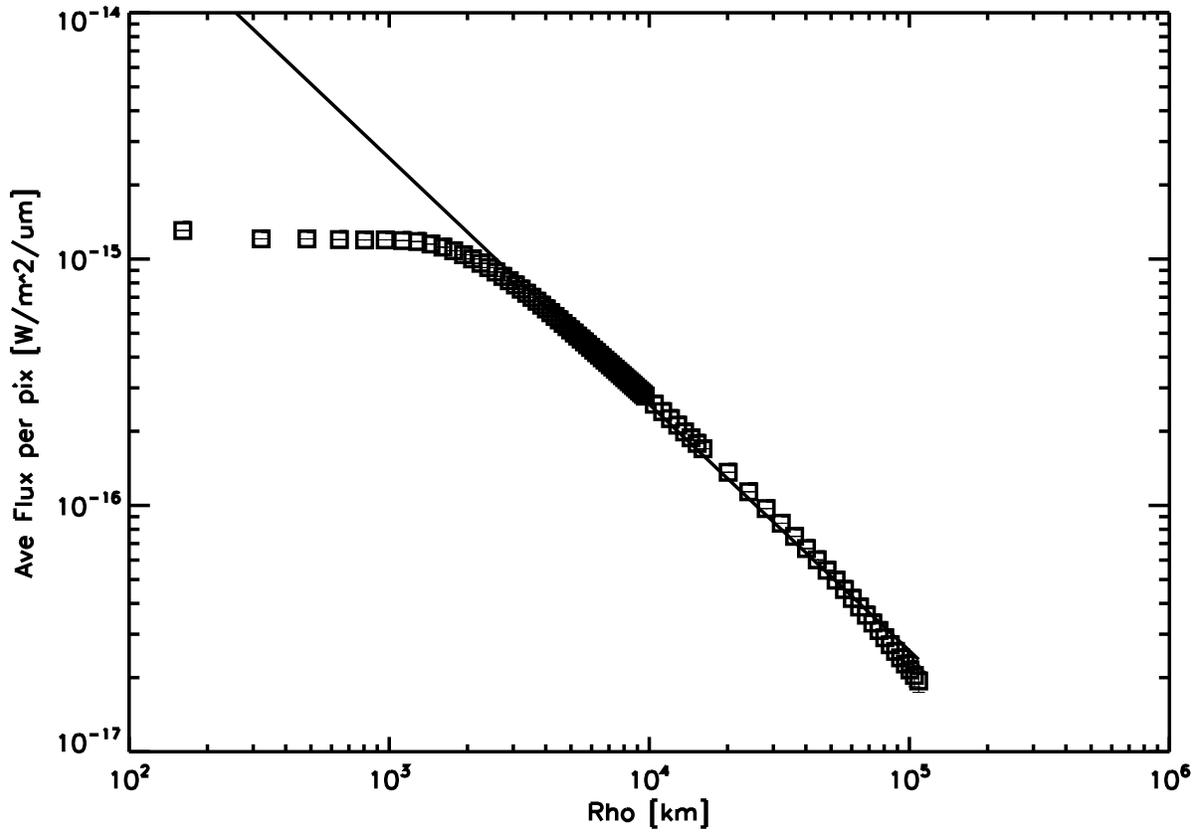}
\caption{Radial profile of fluxes as a function of linear radius as 
measured in $I$ band from the photocenter of comet C/2007 N3 (Lulin) 
obtained on 2009 March 03~UT. Solid line denotes a $1/\rho$ profile.
\label{fig:iradial_prof}}
\end{figure}

The radial profile of comet C/2007 N3 (Lulin) was plotted to assess the 
quality of the data for calculating a dust production rate. The radial 
profile of C/2007 N3 (Lulin) in the V band shows a deviation from the 
conical $1/\rho$ profile \citep{ger92}, suggesting contamination from gas 
such as $C_{2}$ at 5141 \AA. We therefore only calculate the dust 
production in $I$ band. The $I$ band radial profile of C/2007 N3 (Lulin)
is shown in Fig.~\ref{fig:iradial_prof}.

\begin{figure}
\epsscale{1.0}
\plotone{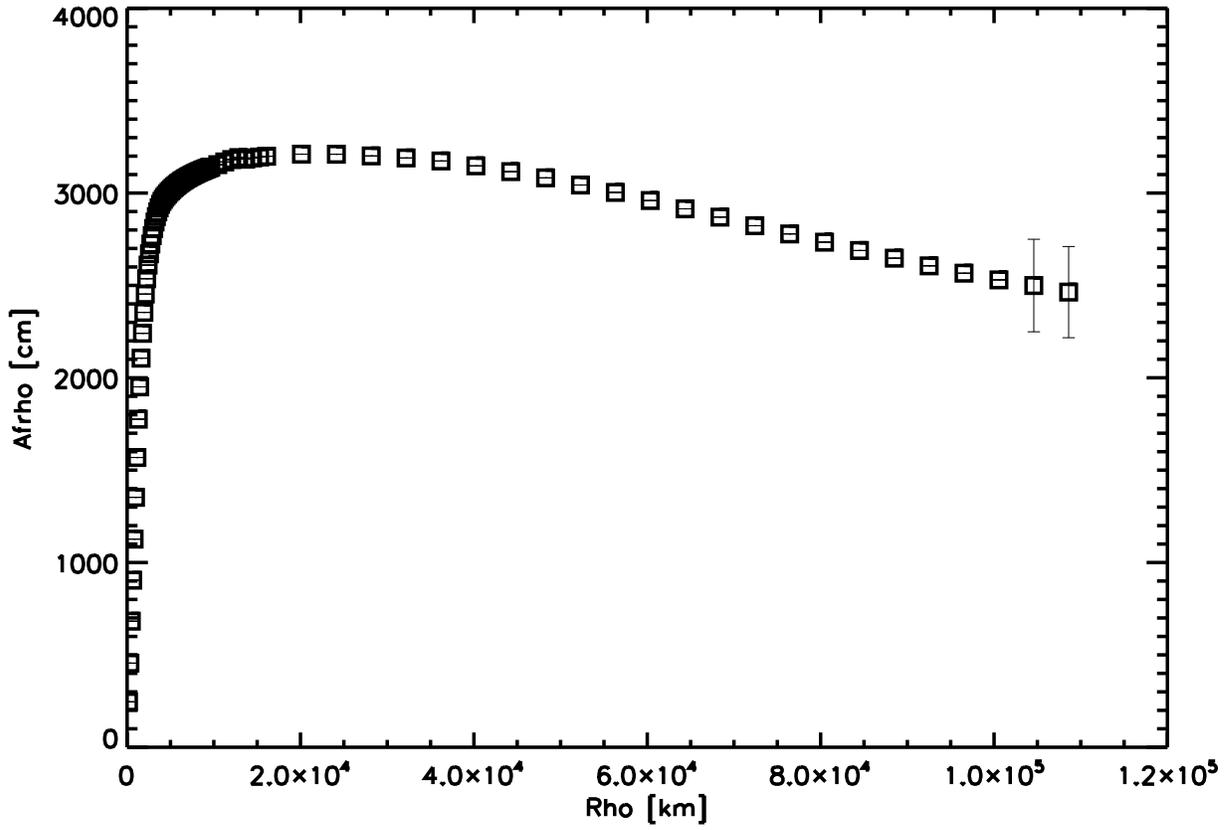}
\caption{$Af \rho$ as a function of $\rho$ derived from $I$ band 
observations of comet C/2007 N3 (Lulin) on 2009 March 03~UT.
\label{fig:afrhoIband}}
\end{figure}

To estimate the rate of dust production in comet C/2007 N3 (Lulin), we 
utilize the $Af\rho$ quantity introduced by \citet{Ahearn84}. This 
quantity serves as a proxy for dust production and when the cometary coma 
is in steady state, the value for $Af \rho$ is an aperture independent 
parameter,

\begin{equation}
Af \rho = \frac{4  \; r_{h}^{2} \;  \Delta^{2}  \; 10^{-0.4(m_{\odot} - m_{comet})}}{\rho}  [cm]
\end{equation}

\noindent where $A$ is the Bond Albedo, $f$ is the filling factor of the 
coma, $m_{\odot}$ is the apparent solar magnitude, $m_{comet}$ is the 
measured cometary magnitude, $\rho$ is the linear radius of the aperture 
at the comet's position (cm)  and $r_{h}$ and $\Delta$ are the 
heliocentric and geocentric distances measured in AU and cm, respectively. 
Cometary magnitudes are observed to follow similar phase angle effects as 
asteroids; therefore we also apply the phase angle correction of 
$m_{comet}(\alpha \; = 0) \; = m_{comet} \; (\alpha) - C \alpha$ where 
$\alpha$ is the phase angle in degrees and C is correction factor of 0.03 
magnitudes per degree, the mean of the correction factors derived by 
\citet{Meech87}. Figure~\ref{fig:afrhoIband} illustrates the progression 
of $Af \rho$ as a function of $\rho$ and Table~\ref{tab:afr_tab}
reports values of $Af \rho$ at a selection of distances from the comet 
photocenter. 

\subsection{\textit{NIR and Polarimetry}} \label{sec:dpol}

Comet C/2007 N3 (Lulin) was sufficiently distant at the epoch of our 
observations that we did not have the spatial resolution to map changes in 
polarization across the coma. Thus it is unknown whether or not there was 
any significant variation in polarization with coma morphology. Comet 
C/1995~O1 (Hale-Bopp), for example, had a high surface brightness jet that 
showed distinctly greater fractional polarization in the NIR than the rest 
of the coma \citep{Jon00}. The only other comet with good spatial 
resolution and polarimetry at NIR wavelengths is comet 
73P/Schwassmann-Wachmann 3 (SW-3), an ecliptic-family comet
\citep{Jon08}. SW-3 passed 
sufficiently close to the Earth to allow us to measure the surface 
brightness and polarization of the coma with a resolution of $40$~km 
\citep{Jon08}. While \citet{Jon08} and \citet{harker11} found strong 
evidence for significant breakup of the dust aggregates released from the 
nucleus of SW-3 taking place over distances of 40-400 km from the nucleus, 
only small changes in the fractional polarization across the inner coma 
were observed. All of the polarization observations of SW-3 were at phase 
angles $> 20^{\circ}$ and showed weak red polarimetric color. Despite the 
strong evidence for significant breakup of dust aggregates, the 
polarization of SW-3 is not distinguishable from other comets. 

\begin{figure}
\epsscale{1.00}
\plotone{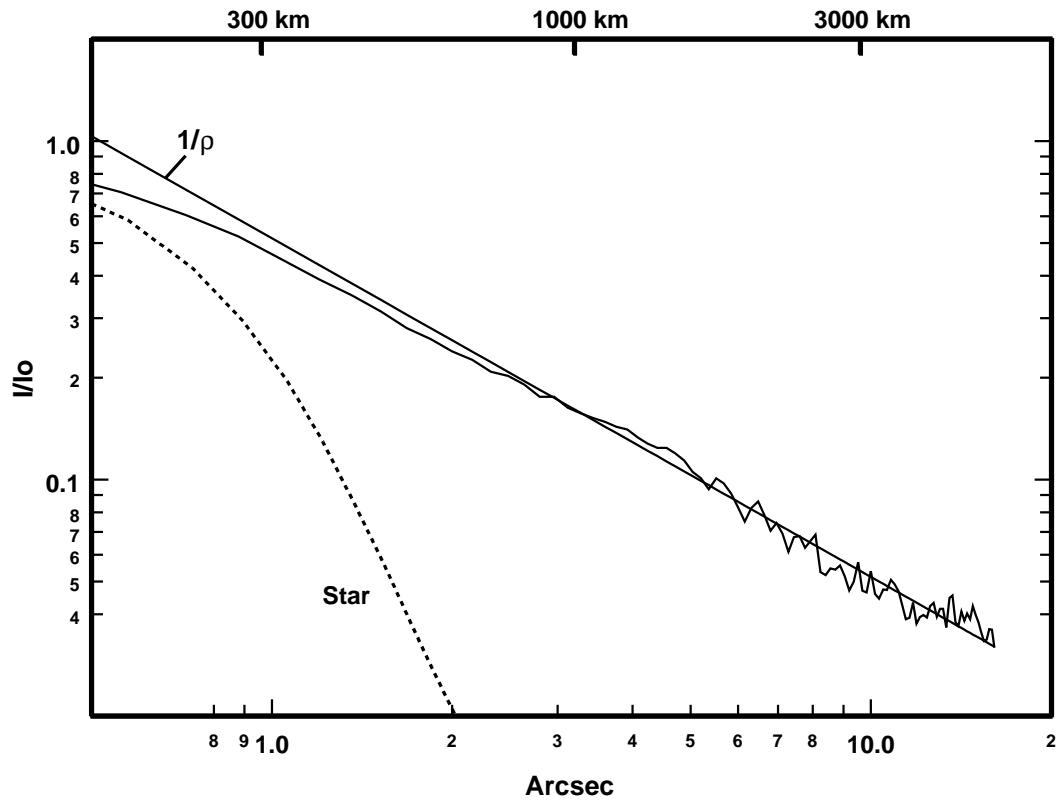}
\caption{An E-W cut across comet C/2007 N3 (Lulin) in 
the $H$ (1.65~\micron) band on 2009 February 27~UT. The profile 
of a star is shown as a 
dotted line. The solid line is a $1/\rho$ fit to the comet coma profile.
\label{fig:EWcut}}
\end{figure}

The surface brightness profile of C/2007 N3 (Lulin) at $H (1.65~\micron)$ 
is shown in Figure \ref{fig:EWcut} along with the profile of a star for 
comparison. Given the seeing of $1\farcs4$ full-width 
half-maximum (FWHM), the 
surface brightness of C/2007 N3 (Lulin) is entirely consistent with a 
simple $1/\rho$ dependence as expected for a constant velocity outflow 
\citep{ger92}. The polarimetry was not of sufficient signal-noise to 
determine if there were any variations in fractional polarization at large 
distances ($> 1000$~km or $3\arcsec$) from the center. 

\begin{figure}
\epsscale{1.10}
\plotone{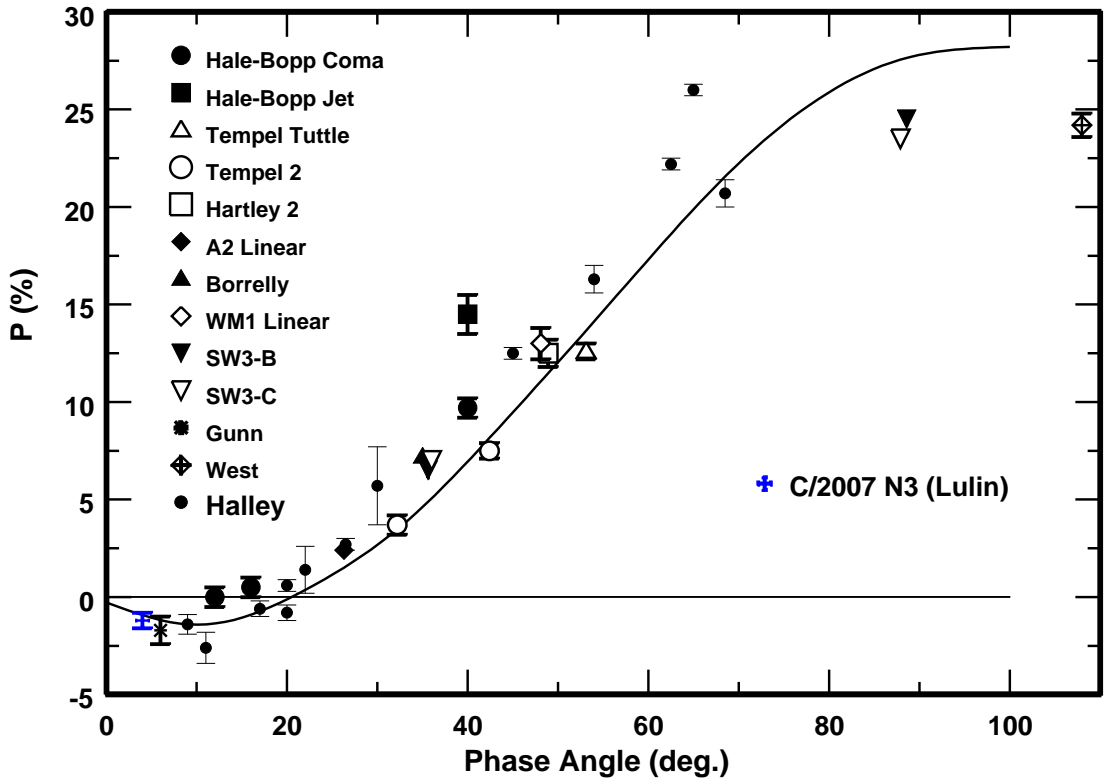}
\caption{Compilation of comet polarimetry in the near-infrared 
\citep[adopted from][]{kel04}. The solid line is the average trend for 
comets in the visual $R$ (0.676~\micron) band \citep{Lev96}. The $H$
(1.65~\micron) measurement for
C/2007~N3~(Lulin) from this work is inset as the blue symbol.
\label{fig:polall}}
\end{figure}

There is very little NIR polarimetry of comets at phase angles 
$< 20^{\circ}$, primarily C/1995 O1 (Hale-Bopp) \citep{Jon00}, 1P/Halley 
\citep{bro87}, and a single observation of 65P/Gunn \citep{kel05}. A 
compilation of NIR polarimetry of comets is shown in 
Fig.~\ref{fig:polall}. The solid line in this figure is the mean phase 
dependence of the polarization in the $R$ band found by \citet{Lev96}. The 
majority, but not all, of the NIR polarimetry lies above this line, at 
least out to phase angles of $60\degr$, consistent with the generally red 
polarimetric color of comets. The precise (low formal error) observations 
of comet C/1995 O1 (Hale-Bopp), which contained unusually small dust 
particles \citep{Wil97, Mas01, harker02, woo99}, \textit{shows no negative 
branch}. The lack of a negative branch in the NIR for C/1995 O1 
(Hale-Bopp) can, to first order, be explained by the added polarization of 
light scattered from more 'Rayleigh-like' particles coming from the jet. 
The addition of these smaller particles increases the fractional 
polarization above what would otherwise be a typical polarization vs. phase 
curve for the rest of the coma. 

\begin{figure}
\epsscale{1.00}
\plotone{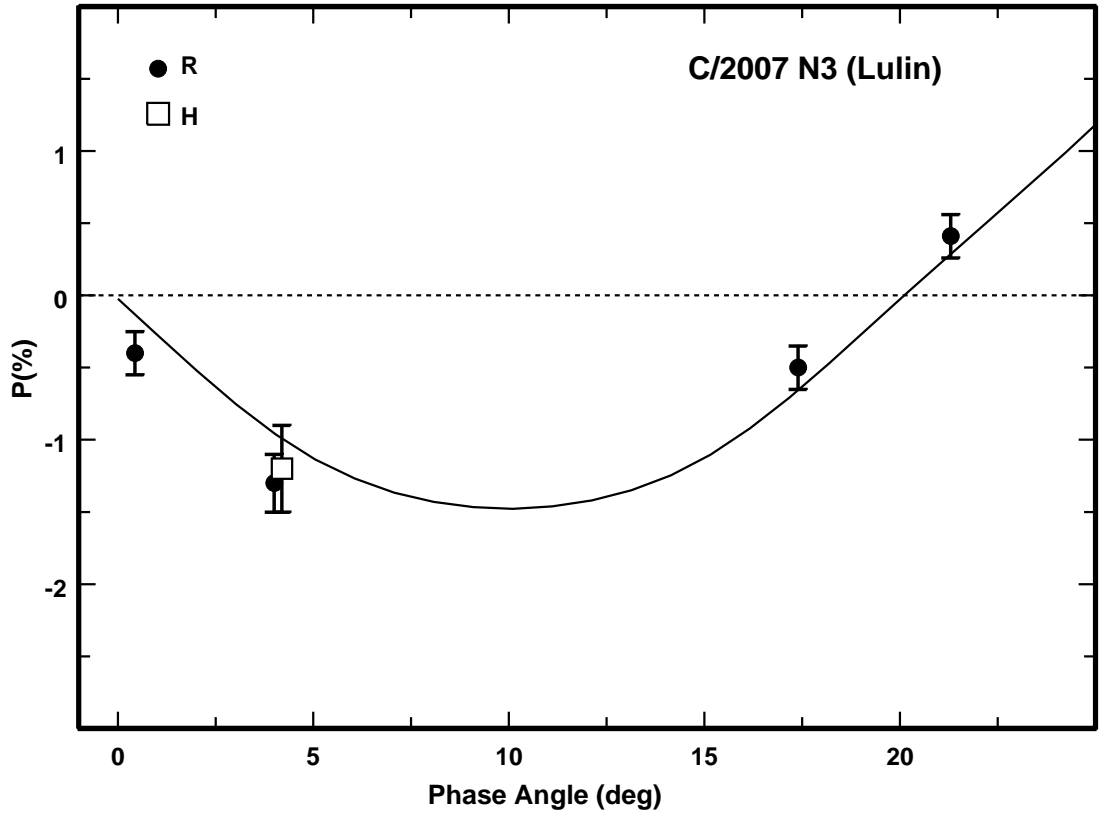}
\caption{Polarization vs. phase angle for C/2007 N3 (Lulin) from data in 
Table~\ref{tab:pol_tab}. The solid line is the average behavior for 
comets in the visual $R$ (0.676~\micron) band \citep{Lev96}. The polarimetry of 
C/2007 N3 (Lulin) is entirely consistent with this trend both 
at $R$ and at $H (1.65~\micron)$.
\label{fig:pol-lulin}}
\end{figure}

Our $H$ band polarimetry of comet C/2007 N3 (Lulin) is the first 
observation of the polarization of a comet in the NIR at a phase angle 
within a few degrees of zero. The polarimetry for comet C/2007 N3 (Lulin) 
given in Table~\ref{tab:pol_tab} is plotted vs. phase angle 
in Fig.~\ref{fig:pol-lulin}. The solid line is the typical dependence 
of polarization on phase angle in the $R$ band mentioned earlier. Our 
$R$ band observations of comet C/2007 N3 (Lulin) are entirely 
consistent with the typical optical behavior of polarization comets. 
The NIR polarization of comet C/2007 N3 (Lulin) is also entirely 
consistent with the contemporaneous visual polarization measurements, 
and the NIR negative polarization is clearly comparable with a 
typical optical negative branch polarization behavior observed in 
most comets at small ($\le 25$\degr) phase angles.  

\subsection{\textit{The 10~\micron{} Silicate Feature}}\label{sec:10um_silicates}

Within the $10~\micron$ spectral range covered by the \textit{Spitzer} 
SL1 module, the observed SED of comets often 
are comprised of amorphous carbon grains, which produce the
underlying featureless emission (continuum) in the 
$8 - 13$~\micron{} wavelength region, and small ($\ltsimeq$ 1~\micron) 
siliceous dust grains which produce broad 
features and distinct resonances in excess of the continuum \citep[for a 
reviews see][]{han_zol10, mskdhw09}. Amorphous silicates with chemical 
composition (stiochiometry) similar to olivine 
(Mg$_{y}$,Fe$_{(1-y)}$)$_{2}$SiO$_{4}$ and pyroxene 
(Mg$_{x}$,Fe$_{(1-x)}$)SiO$_{3}$ with $x=y=0.5$ (Mg/(Mg+Fe)~$=0.5$) 
reproduce the broad width of the 10~\micron{} feature.  The distinct 9.3 
and 10.5~\micron{} emission features are attributed to Mg-rich 
orthopyroxene \citep{woo99,harker02}, while Mg-rich crystalline olivine is 
uniquely identified through its distinct, relatively narrow 11.2~\micron{} 
silicate feature \citep{hanner94}. Mg-rich crystalline species are defined 
as grains with $0.9 \le x \simeq y \ltsimeq 1.0$ \citep{woo08, koike03, 
chihara02}. Frequently in comets, the Mg content of crystalline silicates 
is often significantly larger than that of the amorphous silicate grains. 

\begin{figure}
\epsscale{0.5}
\plotone{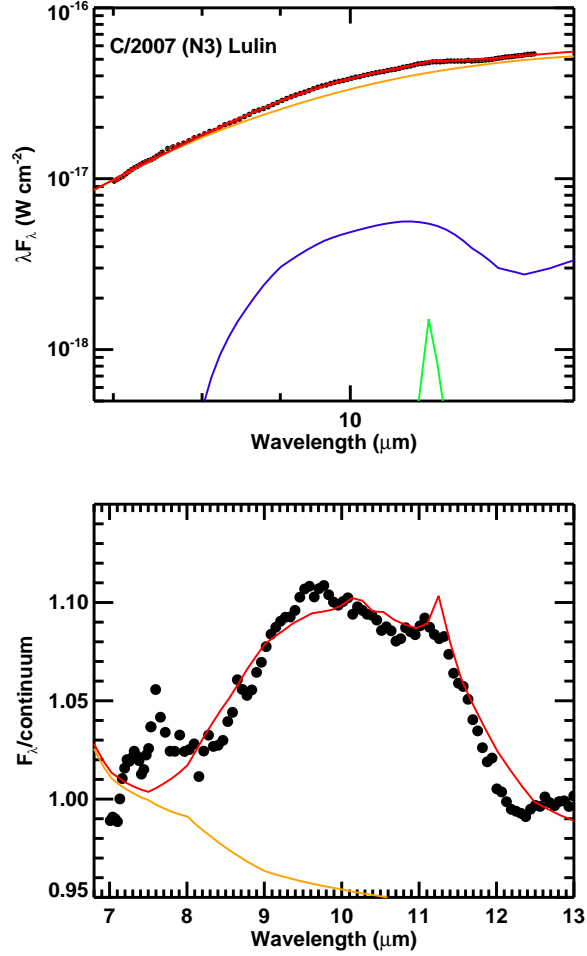}
\caption{$Spitzer$ IRS 7-13~\micron{} spectrum of comet 
C/2007 N3 (Lulin) obtained on 2008 October 04.3~UT ($r_{h} = 1.90$~AU;
$Spitzer$-comet distance $= 1.67$~AU) derived from an 
extraction aperture of $9.25^{\prime\prime} \times 9.25^{\prime\prime}$
centered on the nucleus. Data are shown in black dots. (top) The 
composite thermal model best-fit spectral energy distribution (in
$\lambda  \rm{F}_\lambda$ vs. $\lambda$ space) is indicated
by the solid red line, while the contribution from amorphous pyroxene
($blue$), amorphous carbon ($orange$), and crystalline olivine
($green$) are indicated. (bottom) The observed IRS flux density
divided by a $\simeq 228$~K blackbody continuum 
(F$_{\lambda}$/F$_{\lambda,\, T}$) to highlight spectral details 
of the 10~\micron{} silicate feature. The `feature' near 7.5~\micron{}
is an artifact of IRS order overlap.
\label{fig:spitz_lulin}}
\end{figure}

Weak $10~\micron$ silicate emission is evident in the IRS spectra of comet 
C/2007 N3 (Lulin) (Fig.~\ref{fig:spitz_lulin}). At $10.5$~\micron{} the 
silicate emission is $8.4 \pm 0.1$\% above 
a blackbody curve fit to continuum 
points around $8~\micron$ and $12.5~\micron$. The best fit blackbody has a 
temperature of $\simeq 228$~K. The silicate feature strength at 
$10.5~\micron$ of C/2007 N3 (Lulin) is relatively weak compared to other 
comets, most ecliptic-family comets have $\geq 15$\% silicate feature
strengths \citep{mskdhw09, sitko04}. 

The $10~\micron$ SED of comet C/2007 N3 (Lulin) was modeled using the 
\citet{harker02} thermal emission dust code which assumes that a 
collection of optically-thin, discrete (singular mineralogy) dust 
particles reside at the heliocentric ($r_{h}$) and geocentric ($\Delta$) 
distance of the comet at the epoch of observations (2008~Oct~04.3~UT) 
and adopts a Hanner grain-size distribution \citep[HSGD;][]{hanner83} 
for $n(a)~da$. The $10~\micron$ mineralogy used in the model is derived 
from laboratory studies of interplanetary dust particles 
\citep{woo00}, micrometeorites \citep{bradley99}, the NASA 
\textit{Stardust} mission \citep{brownlee06} as well as other grain 
species employed in other remote sensing thermal models 
\citep[e.g.,][]{hanner94, wooden04}. Additional details of the our 
thermal modeling specifically of comet C/2007 N3~(Lulin) is 
discussed in \citet{woo11}.

The best-fit thermal models (Fig.~\ref{fig:spitz_lulin}, and given in 
Table~\ref{tab:bfmods_tab}) suggest that the silicates in 
C/2007 N3 (Lulin) are dominated by 
pyroxene grains, and include minor amounts of crystalline olivine and 
orthopyroxene. Grains in the coma of C/2007 N3 (Lulin) are relatively 
large, the HGSD peaking at $a_{p}=0.9~\micron$, and moderately porous 
(fractal porosity parameter $D = 2.73$) with a large grain slope $N=4.2$. 
The submicron sized silicate-to-carbon ratio inferred from our models is 
$0.48 \pm 0.06$. The error bars on parameters derived from the
thermal modeling reflect a 95\% confidence limit. We also find 
that the silicate crystalline mass fraction for the submicron to 
micron-size portion of the grain size 
distribution \citep{harker02,moreno03}, defined as $f_{cryst} \equiv$ 
(crystalline)/(crystalline~+~amorphous), for comet C/2007 N3 (Lulin) is 
$0.14 \pm 0.04$. In contrast, other NIC comets such as C/1995~O1 
(Hale-Bopp) which had a very strong 10~\micron{} silicate feature had coma 
dust grains that were fractally very porous ($D = 2.5$) with the HGSD 
peaking at $a_{p}=0.2$~\micron{} \citep{harker02} and mineralogically 
diverse, including amorphous and crystalline forms of both olivine and 
pyroxene with $f_{crys} = 0.68$ \citep{harker02}. The dynamically new 
comet C/2001~Q4 (NEAT) which exhibited a modest $10~\micron$ silicate 
emission had dust grains that were `solid' ($D = 3.0$) with a HGSD peaking 
at $a_{p}=0.3~\micron$ and a with $f_{cryst} = 0.71$ \citep{wooden04}. 

\begin{figure}
\epsscale{1.00}
\plotone{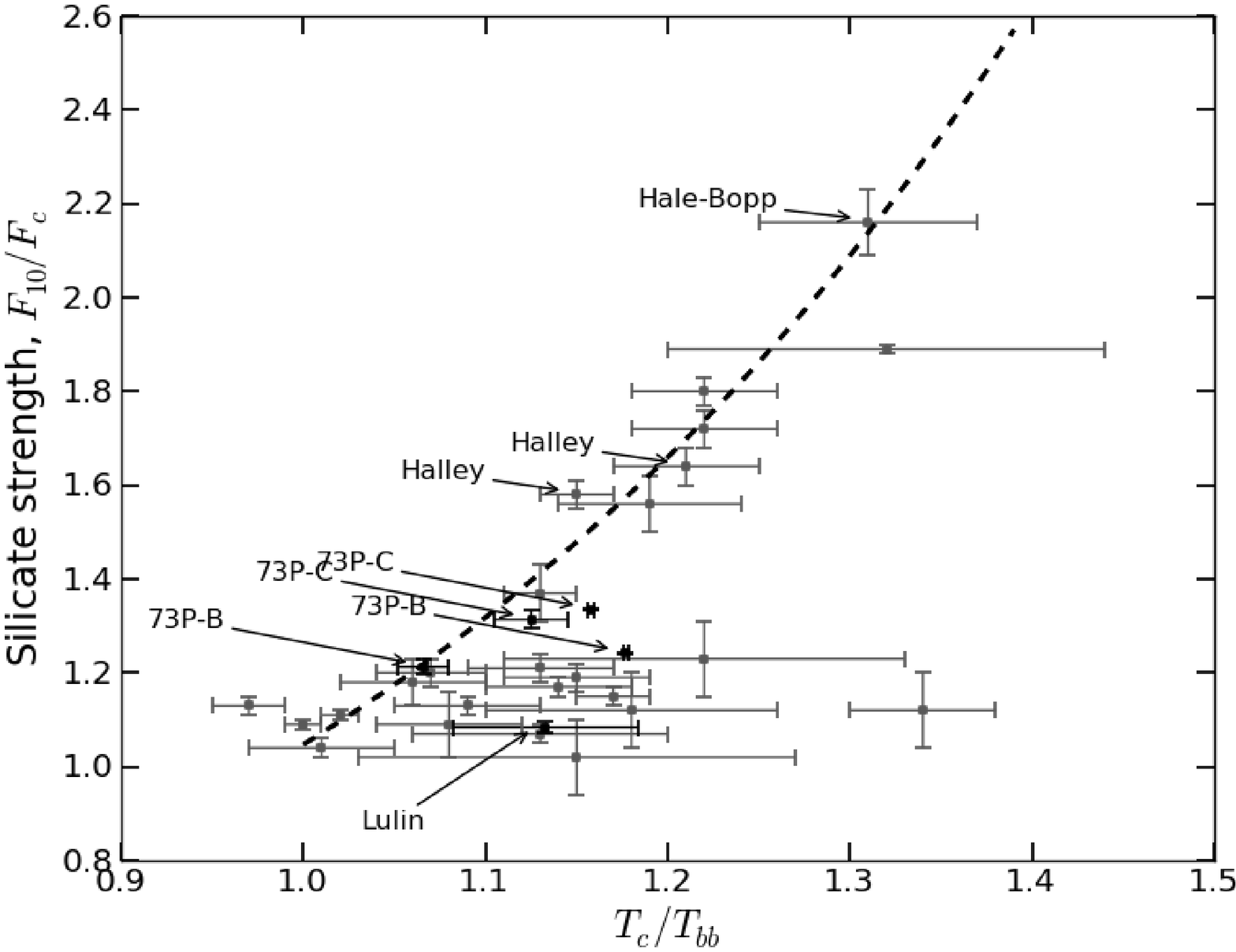}
\caption{The strength of the 10~\micron{} silicate emission feature,
$F_{10} / F_c$, as a function of the excess color temperature above
the equivalent blackbody equilibrium temperature at the Sun-comet
distance, $T_c / T_{BB}$.  The grey points are all comets listed by
\citet{sitko04}.  The individually labeled are comets C/2007 N3 (Lulin)
from this work, 73P/Schwassman-Wachmann fragments B and C
\citep{harker11,sitko11}, 1P/Halley, 
and C/1995 O1 (Hale-Bopp) observed at $r_{h} = 2.73$~AU
\citep{sitko04}. These 5 latter comets all have measured mid-IR 
spectra and near-IR polarization.  The trend-line is the same as 
presented in \citet{sitko04}.
\label{fig:10excess}}
\end{figure}

In Fig.~\ref{fig:10excess} we plot the strength of the 
10~\micron{} silicate emission above the continuum, $F_{10}/F_{c}$,
where $F_{10}$ is integrated feature flux over a bandwidth from
$10-11$~\micron{} and $F_{c}$ is that of the local continuum
at 10.5~\micron, vs. the color temperature excess above 
black body equilibrium at the distance of C/2007 N3 (Lulin), and various
other comets, from the Sun \citep[e.g.,][]{sitko04}. The 
color temperature excess is defined as: 

\begin{equation}
\left( {\frac{{T_{Fit} }}{{T_{BB} }}} \right) = \left( \frac{{T_{Fit} }}{{ {278/\sqrt r_h} }} \right)
\end{equation}

\noindent where $T_{Fit}$ is the black body fit to the observed continuum 
and $r_{h}$ is in AU. For comet C/2007 N3 (Lulin) this temperature 
excess is $1.133 \pm 0.051$, while $F_{10}/F_{c} = 1.084 \pm 0.011$. 
Comet C/1995~O1 (Hale-Bopp), which had a jet 
with much smaller grain aggregate or monomer size dust particles and no 
negative branch in polarization \citep{Jon00}, shows an exceptionally 
strong silicate feature and a large color temperature excess, also 
indicative of small grain aggregates or small grain monomers separated 
from the aggregate \citep{Wil97}. C/2007 N3 (Lulin), in contrast, has a 
very weak silicate feature with little color temperature excess. The large 
compact aggregates that are required by the polarimetry are consistent 
with the thermal IR measurements that show a very weak silicate 
feature. 

The silicate feature strength and continuum temperature excess of 
fragments B and C of comet 73P/Schwassmann-Wachmann observed by 
Gemini \citep{harker11} and \textit{Spitzer} \citep{sitko11} are 
also included in Fig.~\ref{fig:10excess} for completeness.  For 
fragment B, the silicate feature strengths and temperature excesses 
were $1.213 \pm 0.015$ and $1.066 \pm 0.014$ (Gemini), and 
$1.242 \pm 0.002$ and $1.177 \pm 0.001$
(\textit{Spitzer}) respectively.  For fragment C, the 
silicate feature strengths and temperature excesses were 
$1.314 \pm 0.020$ and $1.125 \pm 0.020$ (Gemini), and
$1.335 \pm 0.003$ and $1.158 \pm 0.002$ (\textit{Spitzer}) respectively.




\subsection{\textit{Aggregate Models and Comet NIR Polarimetry}}
\label{sec:polmods}

The negative polarization branch, where the maximum linear polarization is
of order $-2.0$\% is evident in all comets observed at visual wavelengths
at small, $\ltsimeq 25$\degr{}, phase angles \citep[for a review
see][]{mishchenko10, kolv04}. As we mentioned above,
computer simulations of this characteristic, based on aggregate models,
requires either large aggregates (thousands of monomers) or monomers made
of rather transparent material (silicates). These simulations are usually
based on ballistic aggregation that not only allows fitting the
observational data but are also consistent with the origin and evolution
of cometary dust. For example, \citet{kimur06, kimur03} considered two
types of ballistic aggregates: BPCA (Ballistic Particle Cluster
Aggregates) and BCCA (Ballistic Cluster Cluster Aggregate). These two
types of aggregates differ in their porosity. For an equivalent number
$N$ of constituent monomers of radius $a$, the BPCA aggregates are
\textit{more} compact than BCCA, where the porosity is defined as $P = 1
- N(a/a_{c})^{1/3}$ and the characteristic radius $a_{c} \equiv
\sqrt{5/3}\, a_{g}$, where $a_{g}$ is the gyration radius of the
aggregate \citep{kozasa92}, 
$a_{g} = (1/2N^{2}) \times \mathop{\sum}_{i,j=1}^{N}(r(i)-r(j))^{2}$, 
with $r(i)$ the position of the center of the $i^{th}$ 
monomer. \citet{kol07} and 
\cite{kimur06} describe the
polarimetric and IR properties of comet dust using these two types of
ballistic aggregates, consisting of sub-micron monomers (of radius $a
\simeq 0.1$~\micron) with a Halley-type composition which includes
silicates, amorphous carbon, and organic refractory material. These models
can account for the general behavior of the maximum polarization, the
shape of the polarization curve as a function of phase angle $P(\alpha)$,
and the negative polarization branch. In addition, such models also yield
the low geometric albedos (of order $\simeq 5$\% that is typical for comet
dust) and red photometric colors \citep[normalized reflectivity
gradient;][]{jewittmeech86} that are typical for comet dust, for
example C/2004 Q2 (Machholz) \citep{lin2007} and others discussed
in \citet{kolv04} or \citet{hadamcik09}. Indeed, Fig.~2 of 
\citet{joshi11} indicates that
the colors of comet C/2007~N3~(Lulin) are not blue.  For proper
characterization of the observed low geometric albedos and red photometric
of comets, the values of the refractive index used to describe 
the aggregate ensembles are crucial. \citet{kimur06} demonstrate 
that if the imaginary part, $k$, of the index of refraction becomes 
smaller than 0.4 the color of comet dust becomes blue.

The same ballistic aggregates, BPCA and BCCA, were considered by
\citet{lasue09} to model comet polarization. Mixing them with spheroidal
silicate particles they achieved a good fit to the polarization phase
curve. However, \citet{lasue09} do not provide any photometric
characteristics of the dust, specifically the albedo and photometric 
color. This omission does not permit a critical
examination of their model to assess its validity in the study of 
comet C/2007 N3 (Lulin).  The comprehensive study of 
\citet{kimur06} indicates that one might
expect, due to the rather low absorption properties of the \citet{lasue09}
materials (they considered the refractive index of organics equal to
$1.88 + i\, 0.1$ and of silicates $1.62 + i\, 0.003$), that the
\citet{lasue09} dust would demonstrate rather high albedo and blue color.
The erroneously high albedo and blue dust color is also apparent
in the results of \citet{shen09} which uses aggregates to model comet dust;
a aggregate model first developed by \citet{shen08}. In the latter
work, they consider random ballistic aggregates (BAs) constructing
BAM1 and BAM2 particles describing differing monomer migration after
randomized first contact aggregation. BAs have porosities similar to
BPCAs, yet are less porous than BCCAs. However, the complex refractive
indices of the silicates used in \citep{shen09, shen08} models are
low, $1.72 + i\, 0.029$ and $1.71 + i\, 0.031$ in the visible and
IR respectively which lead to rather high albedo and erroneous
photometric colors. For instance, the \citet{shen09} model produces 
an albedo of order $\simeq 12$\% and \textit{blue} dust color.  
Currently, the model most consistent with
the observational and in situ data is that of \citet{kolkim10}
where, following the findings of \textit{Stardust} mission, a
mixture of aggregated and solid particles was considered. The results
of \citet{kolkim10} modeling are characterized not only by the
correct polarization and photometric properties of the dust, including
albedo and colors, but also by the model
organics-to-silicate ratio and ratio of aggregates to solid particles
that are consistent with the results of in situ studies of comets.

An important constraint in interpreting polarimetric aggregate models
applicable to our study of C/2007~N3~(Lulin) near perihelion is our
10~\micron{} spectra\footnote{The synthetic apertures used
to determine the polarization (see Table~\ref{tab:pol_tab})
are comparable to the \textit{Spitzer} extraction apertures.}  which 
show the comet near the epoch of the
optical/NIR polarimetric observations had a weak silicate feature
(\S~\ref{sec:10um_silicates}). Thermal modeling of the 10~\micron{} SED
suggests that the peak of coma grain size distribution $a_{p}$ is of order
1~\micron{} with a large grain slope indicative of a population tail of
grains are extant whose sizes are $>>$~1.0~\micron{}. These latter grains
are large as well as moderately porous, $P_{thermal} \approx f_{vacuum} =
1-(a_{grain}/a_{o})^{D-3}$ \citep{harker02} where $a_{o}$ =
0.1~\micron{} and $D = 2.73$. Large aggregates
are also necessary to explain the negative polarization in the NIR
and the red color of light scattered by comet C/2007~N3~(Lulin)'s 
dust. Extensive optical photometry of the comet near
opposition \citep{joshi11} also indicate a dominance of grains in the
coma larger than 0.1~\micron. Thus, the results derived from 
the scattered light and the thermal 
emission spectra are consistent, and the porosity of grains from 
our thermal models ($\gtsimeq 70$\% for $a_{grain} = 10$~\micron)
suggests that the coma of comet C/2007~N3~(Lulin)
is composed of grains with properties more like BPCA grains rather 
than BCCA grains.



Negative polarization was not observed in the NIR in comet 
C/1995 O1 (Hale-Bopp) whose dust was characterized by small particles. 
Disappearance of the negative polarization in this instance signals that 
at wavelengths $\ge 1.0$~\micron, the size parameter of the dust particles 
became smaller than the wavelength (Rayleigh particles) for which 
polarization is always positive. This is not the case observed in
comet C/2007 (N3) Lulin. The genesis of negative polarization 
is complex \citep[for a detailed discussion see][]{petrova07}. It results 
from a combination of properties of individual monomers, multiple 
scattering, coherent backscattering and near-field effects. At some 
monomer sizes these latter effects, combined with porosity and 
refractive indices, tend to produce effects that work in the same 
direction increasing the negative polarization, but at other monomer 
sizes they produce opposite effects. This is why one can get 
equal negative polarization for different types of aggregates or a big 
change in negative polarization for a small change of, for example, the 
size parameter of the monomer \citep{kolkim10}. Computation of 
negative polarization is still a modeling challenge deserving future attention. 

In our analysis of comet C/2007 N3 (Lulin), we claim that the 
negative polarization does not contradict conclusions regarding the particle 
properties inferred the red polarization color (from the $R$-band through 
the $H$-band) and those derived from the thermal IR observations that 
indicate the grains are large (i.e., micron-sized or larger) and of low 
porosity (perhaps as compact as an aggregate of submicron particles can be).
Another explanation for the negative polarization observed in the coma of
comet C/2007 N3 (Lulin) is that the dust was composed of transparent
silicate particles with little or no carbon content. However, cometary
dust is usually characterized by a low albedo that presumes a large
content of carbon, and photometrically C/2007 N3 (Lulin) does not show any
peculiarity that would allow us to suppose that its albedo is
significantly different. Also, in the case of abundant silicate particles,
we would expect a noticeable spectral dependence of polarization
\citep{zub09}. In addition, our thermal model has a silicate to carbon
ratio of 0.48, inconsistent with dust dominated by transparent silicates.


\section{CONCLUSIONS} \label{sec:conclusion}


We have found that C/2007 N3 (Lulin) clearly has a typical optical
negative branch in the polarization as well as exhibiting a negative
branch in the near-infrared at low phase angles. When compact
aggregates are larger than the longest wavelength considered
it is known that the depth of the negative branch 
does not depend significantly on the wavelength; furthermore, the 
wavelength dependence becomes less pronounced as the phase angle 
decreases \citep{bel09}. We have also found 
that these large, low-porosity aggregates in comet C/2007 N3 (Lulin) are 
consistent with the thermal infrared measurements that showed a 
very weak silicate feature. 

Our thermal model contains moderately porous grains ($D=2.727$), with a 
peak grain size of $0.9~\micron$. This is qualitatively consistent with 
the large aggregate grains needed to explain the polarimetry if the 
monomers are $\sim 0.1$~\micron{} in size. We conclude that the dust in 
comet C/2007 N3 (Lulin) is dominated by large and compact aggregate 
particles, made up of thousands of small monomers. Compact aggregates are 
typical for old periodic comets \citep{kol07} but not for new ones, which 
tend to have more porous aggregates. However, \citet{sitko04} show
that the dust properties of some nearly-isotropic (including
Oort cloud comets) and ecliptic comets 
are similar based on intercomparison of the strength of 
the 10~\micron{} silicate feature versus the color temperature of the dust.
Our polarimetry of comet C/2007 N3 (Lulin), an Oort cloud comet with thermal 
emission properties equivalent to ecliptic-family comets, demonstrate 
that grain structure may account for this 
observation \citep[see Fig.~4 of][]{sitko04}.

The occurrence of a negative branch in the polarization in 
the near-infrared is likely typical for most comets. Comet 
C/1995 O1 (Hale-Bopp), which showed no negative branch in the 
near-infrared, must be considered an anomalous case due to 
the production of significant numbers of very small, submicron grains. 
These latter small coma grains could be individual monomers that have 
broken off from the larger, porous aggregates. For C/2007 N3 (Lulin), the 
more compact aggregates must maintain most of their integrity after 
release from the nucleus. Although aggregate models for the 
polarization and the large porous grain model for the thermal emission of 
comet C/2007 N3 (Lulin) share qualitative features, a rigorous 
quantitative grain/grain-aggregate model that is self consistent for both 
scattering and emission \citep[e.g.,][]{kolkim10} has yet to be 
applied to observations of individual comets. Such an effort 
will be an important future work. 

\section{Acknowledgments}

CEW, ELR, BB, and DEH acknowledge support for this work from the National 
Science Foundation grant AST-0706980.  MSK acknowledges support from NASA 
Planetary Astronomy Grant NNX09AF10G. CEW, DEH, and MSK were also 
supported in part by NASA/JPL \textit{Spitzer} grant 
JPL-01355616. TJJ and MK acknowledge partial support from the 
National Science Foundation grant AST-0937570. The authors also wish
to thank the referee, Dr. D.~H. Wooden, for her insightful review 
that improved the final manuscript.

\clearpage


\clearpage


\begin{deluxetable}{ccccc}
\tablewidth{0pt}
\tablecaption{POLARIMETRY OF COMET C/2007 N3 (Lulin)\tablenotemark{a}\label{tab:pol_tab}}
\tablehead{
\colhead{Date (2009)} & \colhead{UT (Hr)} & \colhead{Phase (deg)}  & \colhead{Filter} & \colhead{P (\%)}
}
\startdata

21 Feb. & 8:00 & 21.3 & $R$ & $0.41 \pm 0.15$ \\
22 Feb. & 7:30 & 17.4 & $R$ & $-0.50 \pm 0.15$ \\
26 Feb. & 10:00 & 0.44 & $R$ &$ -0.40 \pm 0.15$ \\
27 Feb. & 7:30 & 4.0 & $R$ & $-1.30 \pm 0.15$ \\
27 Feb. & 7:30 & 4.0 & $H$ & $-1.20 \pm 0.15$ \\

\enddata
\tablenotetext{a}{$R$-band values determined using a 12\arcsec{} 
diameter synthetic circular aperture, while the $H$-band value is 
measured in a $7\farcs5$ diameter synthetic circular aperture.}
\end{deluxetable}


\begin{deluxetable}{cccc}
\tablewidth{0pt}
\tablecaption{Af $\rho$ VALUES FOR COMET C/2007 N3 (Lulin)\label{tab:afr_tab}}
\tablehead{\colhead{Aperture[arcsec]} & \colhead{$\rho$ [km]} & \colhead{$I$ [mag]} & \colhead{Af $\rho$ [cm]}}
\startdata
28.35 & 10000 & 8.49 $\pm$ 0.06 & 3188 $\pm$ 111\\
42.3 & 15000 & 8.05 $\pm$ 0.07   & 3213 $\pm$138 \\
70.2 & 25000 & 7.49 $\pm$ 0.07 & 3228 $\pm$ 138\\
281.25 & 100000& 6.24 $\pm$ 0.07 & 2545 $\pm$ 109 \\
\enddata
\end{deluxetable}


\begin{deluxetable}{lcc}
\tablewidth{0pt}
\tablecaption{BEST-FIT THERMAL MODEL AND DERIVED PARAMETERS\tablenotemark{a}
\label{tab:bfmods_tab}}
\tablehead{
           &                        & Sub-\micron{} \\
  Dust component & $N_{p} \times 10^{16}$\, \tablenotemark{b} & mass fraction
}
\startdata
Amorphous pyroxene    & $1092 ^{ + 66 }_{ - 46 }$ & $0.17 \pm 0.01$ \\
Amorphous olivine     & $98 ^{ + 24 }_{ - 33 }$   & $0.015 \pm 0.005$ \\
Amorphous carbon      & $5814 ^{ +  7 }_{ -  8 }$ & $0.68 \pm 0.03$ \\
Crystalline olivine   & $268 ^{ + 70 }_{ - 65 }$  & $0.07 \pm 0.02$ \\
Crystalline pyroxene  & $248 ^{ +145 }_{ -156 }$  & $0.07 \pm 0.04$ \\
\cutinhead{Other model parameters}
$\chi^2_\nu$            & 47.2 \\
Degrees of freedom    & 101 \\
Total coma mass       & $(1.096 \pm 0.048) \times 10^{5}$ kg \\
Silicate / carbon     & $0.48 \pm 0.06$ \\

\enddata
\tablenotetext{a}{Uncertainties represent the 95\% confidence level.}
\tablenotetext{b}{Number of grains at the peak of the grain size distribution.}
\end{deluxetable}

%


\clearpage


\clearpage


\clearpage


\clearpage


\clearpage


\clearpage


\clearpage


\end{document}